\documentclass{article}
\usepackage{graphicx} 
\usepackage{amsmath}
\usepackage{bm}
\usepackage{tikz}
\usepackage[numbers]{natbib}
\usepackage{svg}
\usepackage{amsfonts}
\usepackage{graphicx}
\usepackage{caption}
\usepackage{subcaption}
\usepackage{hyperref}
\usepackage{algorithm}
\usepackage{array}
\usepackage{booktabs}
\usepackage{pdflscape}
\usepackage{adjustbox}
\usepackage{authblk}
\usepackage{threeparttable}

\usepackage[margin=3cm]{geometry}
\tikzstyle{arrow1} = [black,thick,->,>=stealth]
\tikzstyle{arrow2} = [red, dashed,thick,->,>=stealth]
\usetikzlibrary{arrows, positioning, bayesnet}

\title{Warfare Ignited Price Contagion Dynamics in Early Modern Europe}
\author[1]{Emile Esmaili}
\author[2]{Michael J. Puma}
\author[3]{ Francis Ludlow}
\author[4]{Poul Holm}
\author[5]{Eva Jobbova}
\affil[1,2]{Center for Climate Systems Research, Columbia University}
\affil[3,4,5]{ Trinity Centre for Environmental Humanities, Trinity College Dublin}
\date{\today}

\begin{document}

\maketitle

\begin{abstract}
Economic historians have long studied market integration and contagion dynamics during periods of warfare and global stress, but there is a lack of model-based evidence on these phenomena. This paper uses an econometric contagion model, the Diebold-Yilmaz framework, to examine the dynamics of economic shocks across European markets in the early modern period. Our findings suggest that key periods of violent conflicts significantly increased food price spillover across cities, causing widespread disruptions across Europe. We also demonstrate the ability of this framework to capture relevant historical dynamics between the main trade centers of the period.
\end{abstract}



\section{Introduction}

Economic historians have consistently emphasized the expansion of markets as a pivotal factor in driving economic growth, particularly in pre-industrial Europe. Market integration, is reflected in the degree to which prices across different regions move together, has been highlighted as a crucial element in this economic progress \cite{kelly1997dynamics,epstein2000freedom,malanima2009pre}. The examination of market integration has a long history in economic research, with quantitative and statistical analyses dating back to the 1950s \cite{achilles1957getreidepreise}, and most recent studies primarily used econometric frameworks to investigate price convergence and comovement across different regions, allowing for a deeper understanding of how integrated markets were over time \cite{federico2011did,federico2012much,federico2024market,bateman2011evolution}. While there is some debate among scholars regarding the precise timeline of market integration, a general consensus exists that this process began in the late fifteenth century, saw rapid advancements up until the early seventeenth century, and then experienced a temporary stall. Following the Thirty Years War, market integration resumed and accelerated significantly after the Napoleonic Wars, largely due to widespread trade liberalization \cite{Federico_Schulze_Volckart_2021,chilosi2013europe,jacks2005intra}.

In our study, we seek to build upon the extensive literature on market integration by shifting the focus from traditional measures of price convergence and comovement to a more dynamic analysis of price contagion and network effects. Many scholars, including Fernand Braudel \cite{braudel1981civilization}, have explicitly studied these networks of trade in early-modern Europe, but to our knowledge, this work is the first to apply an econometric contagion model to study these dynamics. We use the Diebold-Yilmaz (DY) framework, a methodological tool that has been influential in the study of financial and trade networks \cite{DY2009, DY2012, DY2014}. Despite its effectiveness in modeling interconnected systems and quantifying the extent of shock propagation effects (or spillovers), the DY framework has not yet been applied in the field of economic history. This is surprising given its potential to offer new insights into historical market integration, especially when considering the impact of external shocks, such as wars or political upheavals, on economic systems.

The DY framework is particularly appealing because it leverages Vector Autoregressive (VAR) models, which are popular in macroeconomics for capturing the dynamic relationships between multiple time series. VAR models allow estimation of residual covariances among variables. These quantities are essential to qualify the influence that an outside shock on one variable can have on all other variables in the system: an effect defined as the spillover.  Calculating these spillover effects provides a richer understanding of market dynamics beyond measures of price alignment. The DY model is popular in financial economics \cite{DIEBOLD2023115,kakran2023novel} and in research of trade networks \cite{demirer2018estimating}. Furthermore, by extending the VAR model to a dynamic network context, the DY framework enables researchers to visualize and quantify the evolving patterns of economic contagion over time. This makes it a powerful tool for analyzing the interconnectedness and inter-dependencies of historical markets, providing a more nuanced view of how shocks propagated through economic networks.

We believe that the intersection of econometrics and network science offered by the DY model provides a robust framework for exploring historical market integration and price dynamics in new and insightful ways. Our study aims to fill this gap by applying the DY framework to examine the dynamics of price contagion in early modern Europe, thus contributing to an understanding of market integration through a new lens.

\section{Data and Methods}
\subsection{Data}

\begin{figure}[htbp!]
    \centering
    \includegraphics[width=0.85\linewidth]{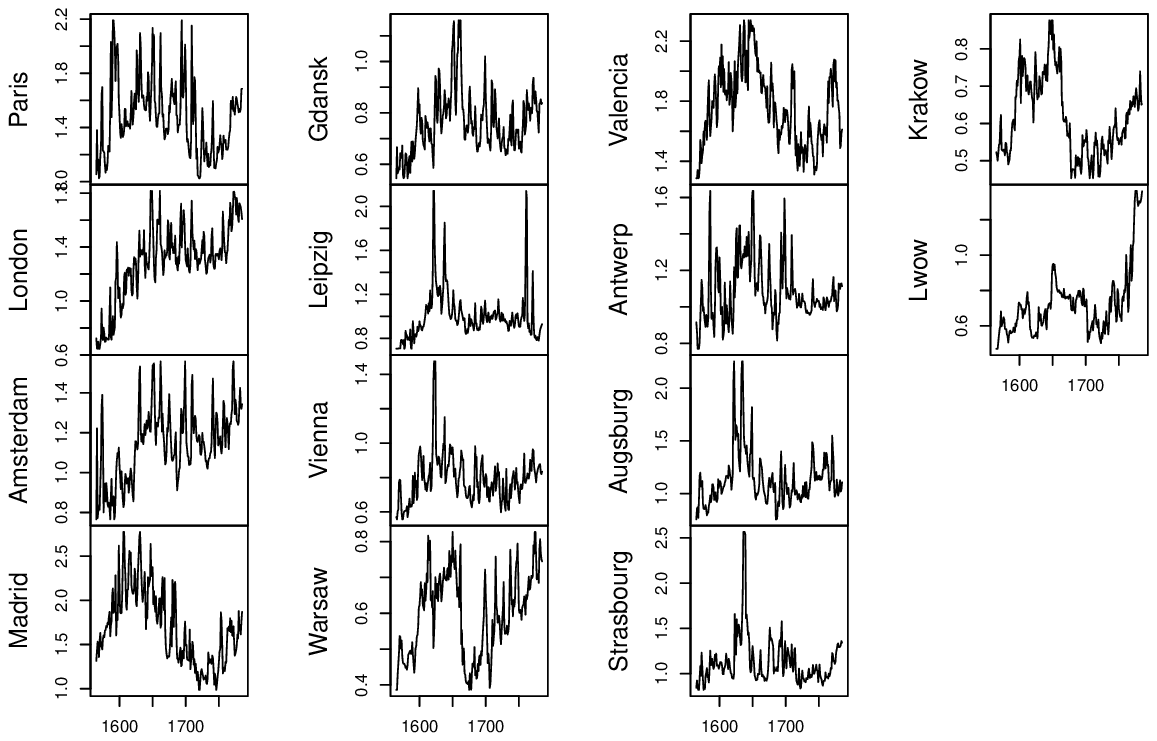}
    \caption{CPI time-series}
    \label{fig:timeseries}
\end{figure}

\begin{figure}[htbp!]
    \centering
    \includegraphics[width=0.5\linewidth]{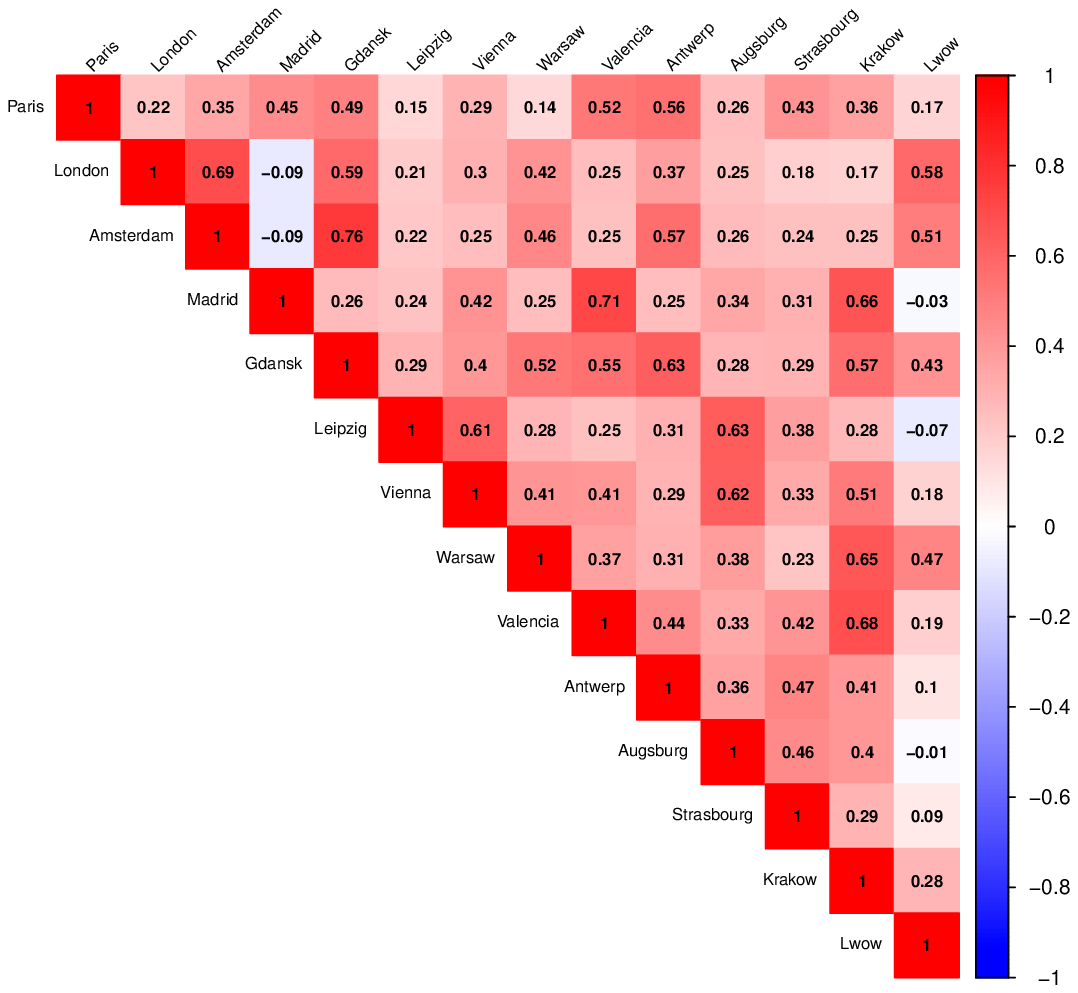}
    \caption{Pearson correlation matrix of prices}
    \label{fig:corr_mat}
\end{figure}

We use yearly Consumer Price Indices (CPI) collected from \citet{allen2001great} for 14 European cities between 1562 to 1793. The cities included in the analysis are: Paris, London, Antwerp, Amsterdam, Gdansk, Lwow, Warsaw, Vienna, Krakow, Madrid, Valencia, Augsburg, Leipzig and Strasbourg. We winsorize the raw CPI values to 1 percent level (location-wise) and plot them in figure \ref{fig:timeseries}. We also display their Pearson correlation matrix in figure \ref{fig:corr_mat}. 

To estimate the number of conflict fatalities, we use the Conflict Catalog, a taxonomy of the most severe recorded violent conflicts \cite{Brecke1999}, and focus on Eastern and Western Europe (regions coded as 3 and 4 in the dataset). To calculate the yearly falities from events spanning multiple years, the total fatalities for each event are evenly distributed across all the years of its duration. For an event \(i\) with total fatalities \(\text{Deaths}_i\), start year \(\text{StartYear}_i\), and end year \(\text{EndYear}_i\), the contribution to each year in the range is given by:
\[
\text{Fatalities Per Year}_i = \frac{\text{Fatalities}_i}{\text{EndYear}_i - \text{StartYear}_i + 1}.
\]

The total fatalities in a given year \(y\) are the sum of contributions from all events that span that year. This is expressed as:
\[
\text{Fatalities in Year } y = \sum_{i=1}^{n} \frac{\text{Fatalities}_i}{\text{EndYear}_i - \text{StartYear}_i + 1} \cdot \textbf{1}_{\{y \in [\text{StartYear}_i, \text{EndYear}_i]\}},
\]
where \(\textbf{1}_{\{y \in [\text{StartYear}_i, \text{EndYear}_i]\}}\) is an indicator function that equals 1 if \(y\) is within the range of the event \([\text{StartYear}_i, \text{EndYear}_i]\), and 0 otherwise, for $n$ events in years $y$.

This method ensures that fatalities are evenly distributed across the years spanned by each event and that overlapping contributions from multiple events are appropriately summed for each year.

\subsection{Methods}

\subsubsection{Contagion model}

As noted, we use the contagion model introduced by Diebold-Yilmaz (DY).\cite{DY2009, DY2012, DY2014}. 
The framework uses a VAR(p) model's Forecast-Error Variance Decomposition (FEVD) to compute the spillover from market $i$ to market $j$. Consider $\mathbf{Y} = (y_1,y_2,...y_n)$ our $N \times T$ dimensional time-space matrix where every $y_i$ is the CPI time-series of length $T$ for location $i$. the VAR(p) model is written as $y_t = \sum_{i=0}^{p}\phi_i y_{t-i} + \epsilon_t$, where $\epsilon \sim N(0,\Sigma)$. 
The FEVD is populated by elements $d^H_{ij}$, which give the proportion of the $H$-step forecast error variance of variable $y_j$ that is driven by an orthogonal shock to $y_i$. 

Denoting the $H$-step-ahead forecast error variance decompositions by $d^H_{ij}$, for $H = 1, 2, \ldots$, we have:

\begin{equation}
d^H_{ij} = \frac{\sigma_{ii}^{-1} \sum_{h=0}^{H-1} (e_i' A_h \Sigma e_j)^2}{\sum_{h=0}^{H-1} (e_i' A_h \Sigma A_h' e_i)},
\end{equation}

where $\Sigma$ is the variance matrix for the error vector $\varepsilon$, $\sigma_{ii}$ is the standard deviation of the error term for the $i$th equation, $e_i$ is the selection vector with one as the $i$th element and zeros otherwise, and $A_h$ is the coefficient matrix of the Moving Average representation of the VAR(p) model, where \[
y_t = \sum_{i=0}^{\infty} A_i \varepsilon_{t-i},
\]
where the $N \times N$ coefficient matrices $A_i$ obey the recursion:
\[
A_i = \Phi_1 A_{i-1} + \Phi_2 A_{i-2} + \cdots + \Phi_p A_{i-p},
\]
with $A_0$ being an $N \times N$ identity matrix and $A_i = 0$ for $i < 0$.

\begin{table}[h!]
\centering
\caption{Diebold-Yilmaz Spillover Table}
\begin{tabular}{ccccc|c} 
\toprule
 & $y_1$ & $y_2$ & $\cdots$ & $y_N$ & \textbf{To Others} \\ \midrule
$y_1$ & $d^H_{11}$ & $d^H_{12}$ & $\cdots$ & $d^H_{1N}$ & $\sum_{j=1}^N d^H_{1j}, \, j \neq 1$ \\
$y_2$ & $d^H_{21}$ & $d^H_{22}$ & $\cdots$ & $d^H_{2N}$ & $\sum_{j=1}^N d^H_{2j}, \, j \neq 2$ \\
$\vdots$ & $\vdots$ & $\vdots$ & $\ddots$ & $\vdots$ & $\vdots$ \\
$y_N$ & $d^H_{N1}$ & $d^H_{N2}$ & $\cdots$ & $d^H_{NN}$ & $\sum_{j=1}^N d^H_{Nj}, \, j \neq N$ \\ \midrule
\textbf{From Others} & $\sum_{i=1}^N d^H_{i1}, \, i \neq 1$ & $\sum_{i=1}^N d^H_{i2}, \, i \neq 2$ & $\cdots$ & $\sum_{i=1}^N d^H_{iN}, \, i \neq N$ & $\frac{1}{N} \sum_{i,j=1}^N d^H_{ij}, \, i \neq j$ \\
\bottomrule
\end{tabular}
\end{table}

Diebold and Yilmaz define $d^H_{ij}$ as a pairwise directed spillover from $i$ to $j$:

\begin{equation}
S^H_{i \rightarrow j} = d^H_{ij}.
\end{equation}

The pairwise spillovers allow for the construction of more aggregated spillover indices. For example, the off-diagonal column sums indicate to what degree the $H$-step forecast error variation of variable $y_j$ is driven by other variables in the system. Diebold and Yilmaz define inward spillovers as:

\begin{equation}
S^H_{j \leftarrow \bullet} = \sum_{i=1}^N d^H_{ij}, \, i \neq j.
\end{equation}

Conversely, the off-diagonal row sums indicate to what degree variable $y_j$ drives the variation of all other variables in the system. Outward spillovers are therefore defined as:

\begin{equation}
S^H_{j \rightarrow \bullet} = \sum_{i=1}^N d^H_{ji}, \, i \neq j.
\end{equation}

Total spillovers in the system are finally defined as the average of inward or outward spillovers:

\begin{equation}
S^H = \frac{1}{N} \sum_{i,j=1}^N d^H_{ij}, \, i \neq j.
\end{equation}

We first test for first-order stationarity (a requirement to meet the assumptions of a VAR model) using the Augmented Dickey Fuller test \cite{dickeyfuller} on first-differenced CPI data. All time-series are first-order stationary at the 1\% significance level (see table \ref{tab:adf_results}). We proceed to construct a dynamic spillover index for early modern Europe, by computing the total spillover dynamically every year with a chosen rolling window. We use a VAR(1) model whose order $p=1$ was chosen by Information criteria on the full sample on first-differenced CPIs. We use a standard 10-year horizon forecast for the FEVD and analyze different choices of rolling window for robustness. We also look at the net spillovers ($S^H_{j \rightarrow \bullet} - S^H_{j \leftarrow \bullet}$) in a dynamic, rolling-window setting. 

\begin{table}[ht]
\centering
\caption{Augmented Dickey-Fuller test statistics and p-values by location. Lag order of 10}
\begin{tabular}{lrl}
  \hline
Location & Statistic & p-value \\ 
  \hline
Antwerp & -6.80 & $<$0.01 \\ 
  Amsterdam & -6.29 & $<$0.01 \\ 
  London & -7.61 & $<$0.01 \\ 
  Paris & -6.51 & $<$0.01 \\ 
  Strasbourg & -5.92 & $<$0.01 \\ 
  Valencia & -5.52 & $<$0.01 \\ 
  Madrid & -6.91 & $<$0.01 \\ 
  Augsburg & -7.85 & $<$0.01 \\ 
  Leipzig & -6.28 & $<$0.01 \\ 
  Vienna & -6.88 & $<$0.01 \\ 
  Gdansk & -6.36 & $<$0.01 \\ 
  Krakow & -4.12 & $<$0.01 \\ 
  Warsaw & -5.09 & $<$0.01 \\ 
  Lwow & -5.83 & $<$0.01 \\ 
   \hline
\end{tabular}
 
\label{tab:adf_results}
\end{table}

As highlighted by \citet{DY2014} the spillover table can be naturally seen as a directed weighted graph adjacency matrix, where the nodes are the cities that we study and the edges the spillover from one to the other. This enables us to construct a time-varying network of food price contagion dynamics across European markets.

\subsubsection{Regression analyses of spillover and conflict}

To study the statistical relationship between the total spillover from the Diebold Yilmaz model and the number of conflict fatalities, we first use an Ordinary Least Squares (OLS) regression that controls for the level of prices:

\[
\text{Spillover Index}_t = \beta_0 + \beta_1 \log(\text{Fatalities}_t) + \beta_2 \text{CPI}_t + \epsilon_t
\]
where $\epsilon$ is a zero-mean Gaussian error term. Spillover Index is the average (over various choices of rolling windows) total spillover index and Fatalities is the number of non-zero conflict fatalities. We report New-West standard errors \cite{newey-west} that are robust to common violations of the linear model hypotheses: autocorrelation and heteroskedasticity of the residuals. 

Next we seek to understand whether the relationship between conflict and price spillovers varies across the distribution of spillovers (i.e. is it more pronounced for higher levels of spillovers?). While ordinary least squares regression focuses on the mean of the response variable at each value of the predictors, one might also want to know more details about the probability distribution of the response variable—for example, how the predictors relate to the occurrence of extreme values. This can be determined by means of quantile regression. We run regressions in key quantiles on the spillover index (25th, 50th, 75th and 90th) following the same specification as the OLS regression, and report boostrapped standard errors.

\section{Results}

\subsection{Conflict and Price Contagion}

\subsubsection{General relationship}

We use the dynamic spillover to assess the evolution of contagion dynamics between the cities we study. For this dynamic analysis, we estimate spillovers from a VAR(1). In this setup we are dealing with a computational issue: we would face a rank-deficiency constraint (i.e. not enough time-series given the number of locations) if we were to move on to higher order VARs. For reference, the original DY paper \cite{DY2009} uses daily data, a window size of 200 for 2000 samples and a VAR(4). We run the VAR(1) with window sizes ranging from 30 to 40 and average over windows to compute the average spillover. 

\begin{figure}[htbp!]
    \centering
    \begin{subfigure}[b]{0.8\linewidth}
        \centering
        \includegraphics[width=\linewidth]{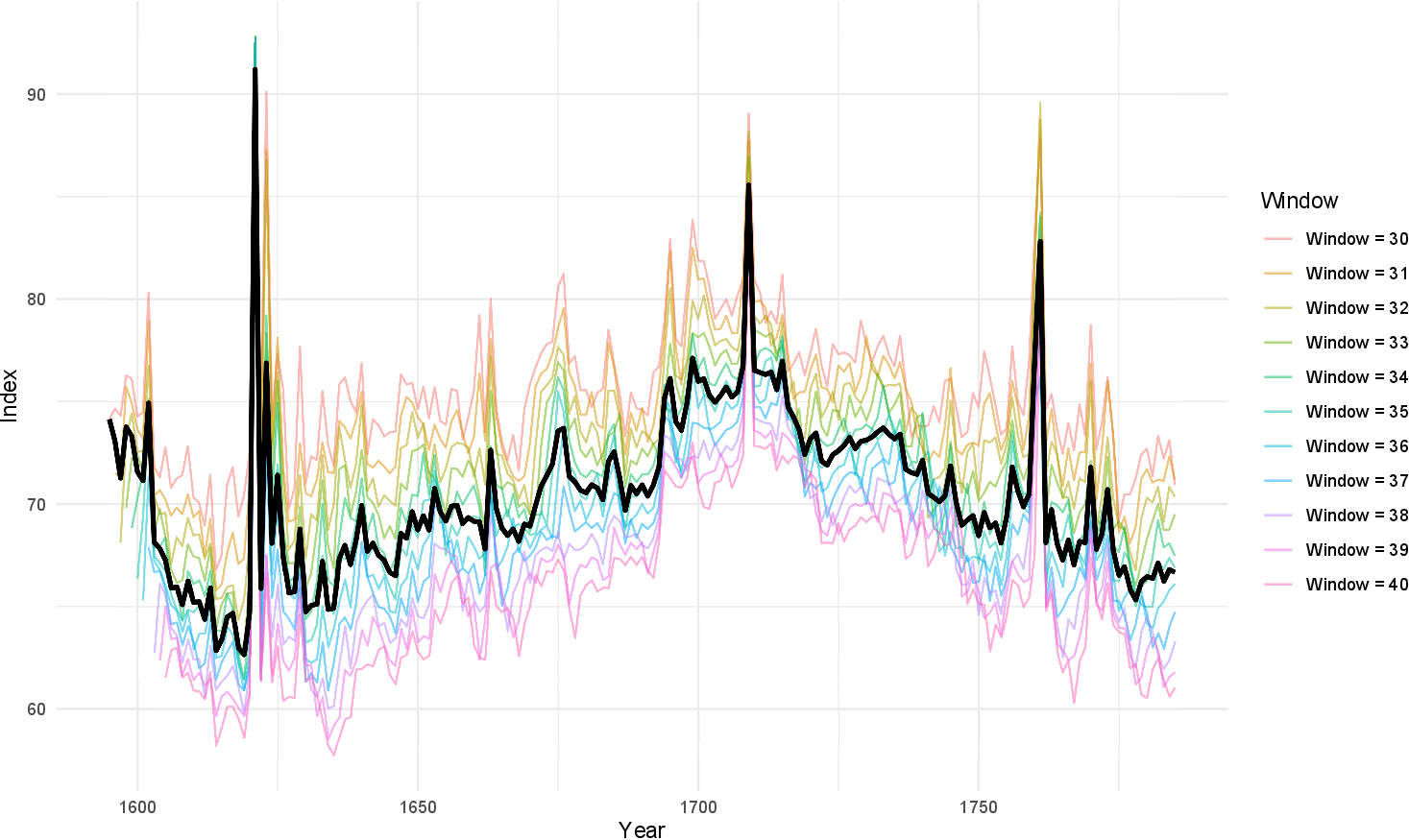}
        \caption{}
        \label{fig:spiloverwindows}
    \end{subfigure}
    \hfill
    \begin{subfigure}[b]{0.8\linewidth}
        \centering
        \includegraphics[width=\linewidth]{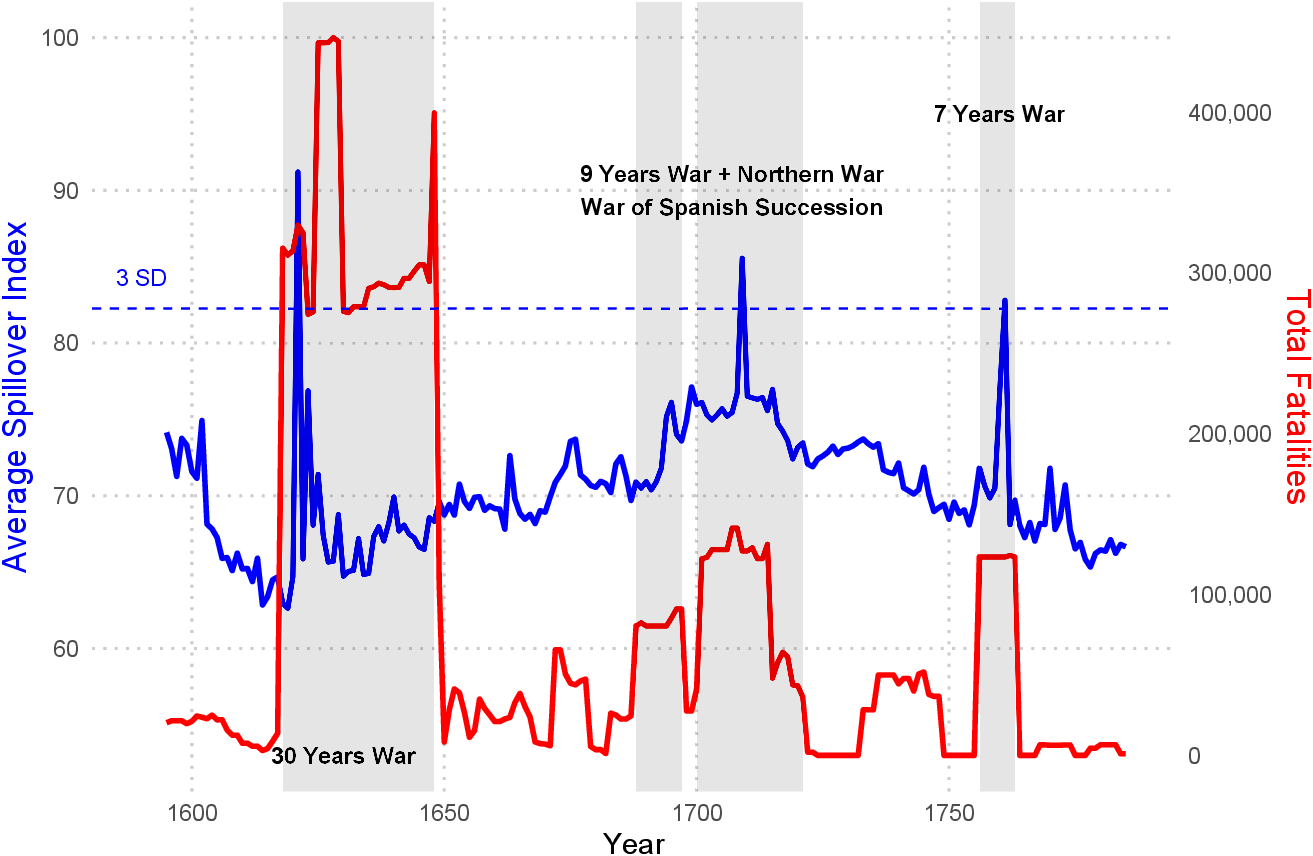}  
        \caption{}
        \label{fig:spilloverdeaths}
    \end{subfigure}
    \caption{ (a) Dynamic total spillover using 10-year forecast horizon, VAR(1), various choices of rolling window (b) Average Total Spillover and Total Conflict Fatalities over Western and Eastern Europe  }
    \label{fig:dynamic_spillover_figs}
\end{figure}

We see a striking spike in spillover during the key warfare period of the 17th and 18th centuries (figure \ref{fig:dynamic_spillover_figs}): the Thirty-Years War (1618-1648), the Nine-Years War (1688-1697), the 1700-1721 period comprising of the Great Northern War (1700-1721) and the War of Spanish Succession (1701-1714) and the Seven Years War (1756-1762). The main spillover spikes occur right after these conflicts begin, and coincide with subsequent periods of elevated conflict fatalities, which is strong evidence to claim that these events propagated price disruptions throughout European markets at large.

To quantify the interaction between conflict severity, proxied by the number of fatalities, and the spillover index, we examine whether the behavior of both time series, as shown in figure \ref{fig:spilloverdeaths}, provides evidence that conflict intensified food price spillovers.

\begin{figure}[htbp!]
    \centering
    \begin{subfigure}[b]{0.49\linewidth}
        \centering
        \includegraphics[width=\linewidth]{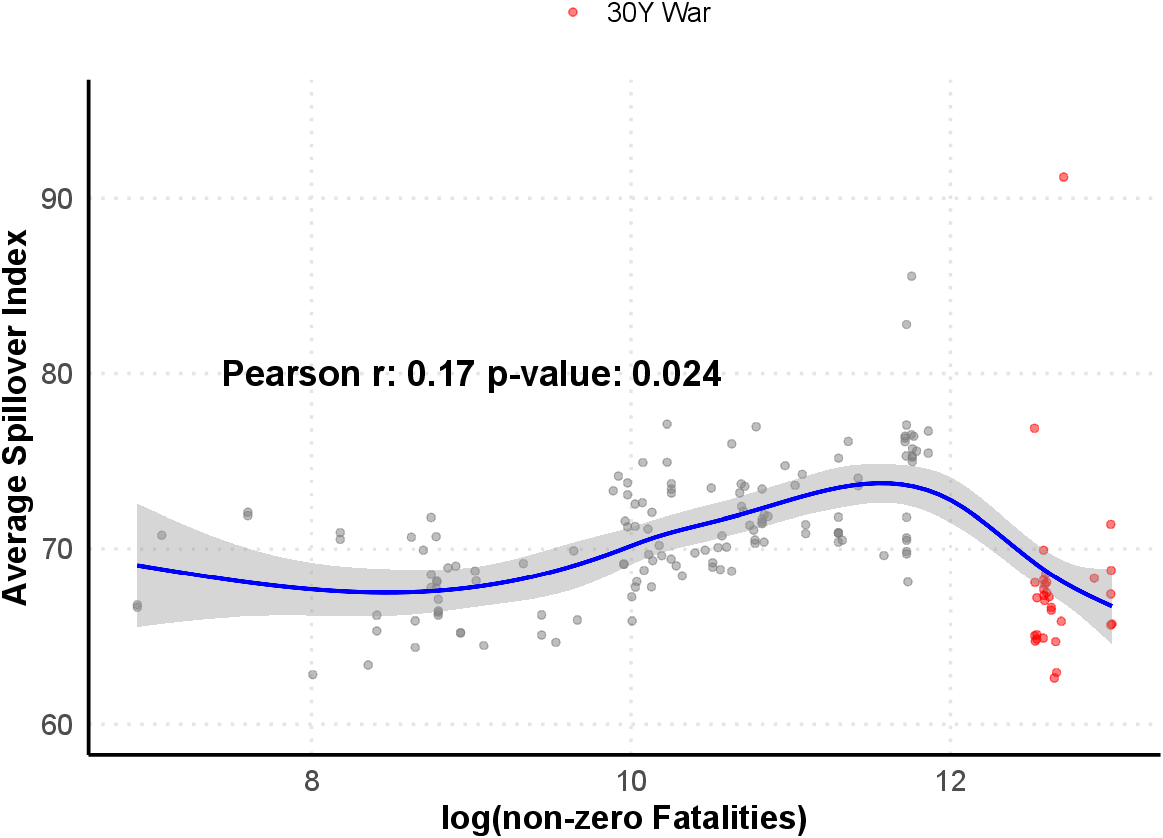}
        \caption{}
        \label{fig:scatterplot}
    \end{subfigure}
    \hfill
    \begin{subfigure}[b]{0.49\linewidth}
        \centering
        \includegraphics[width=\linewidth]{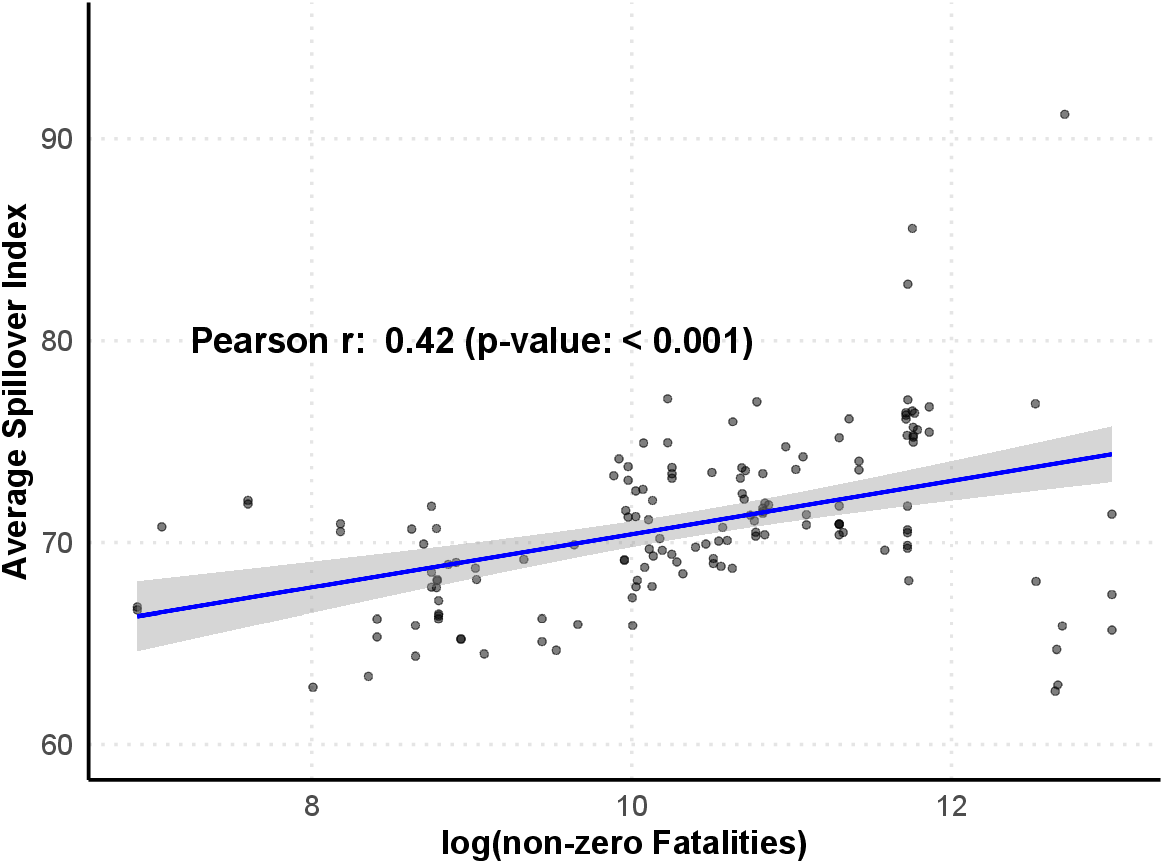}  
        \caption{}
        \label{fig:scatterplot_excl}
    \end{subfigure}
    \caption{Raw data relationship between non zero fatalities (log) and total spillover index (a) Full sample relationship, the fitted curve is a restricted cubic spline (b) Excluding 1628-1648, line fitted with Ordinary Least Squares}
    \label{fig:scatterplots}
\end{figure}

\begin{figure}[htbp!]
    \centering
    \begin{subfigure}[b]{0.49\linewidth}
        \centering
        \includegraphics[width=\linewidth]{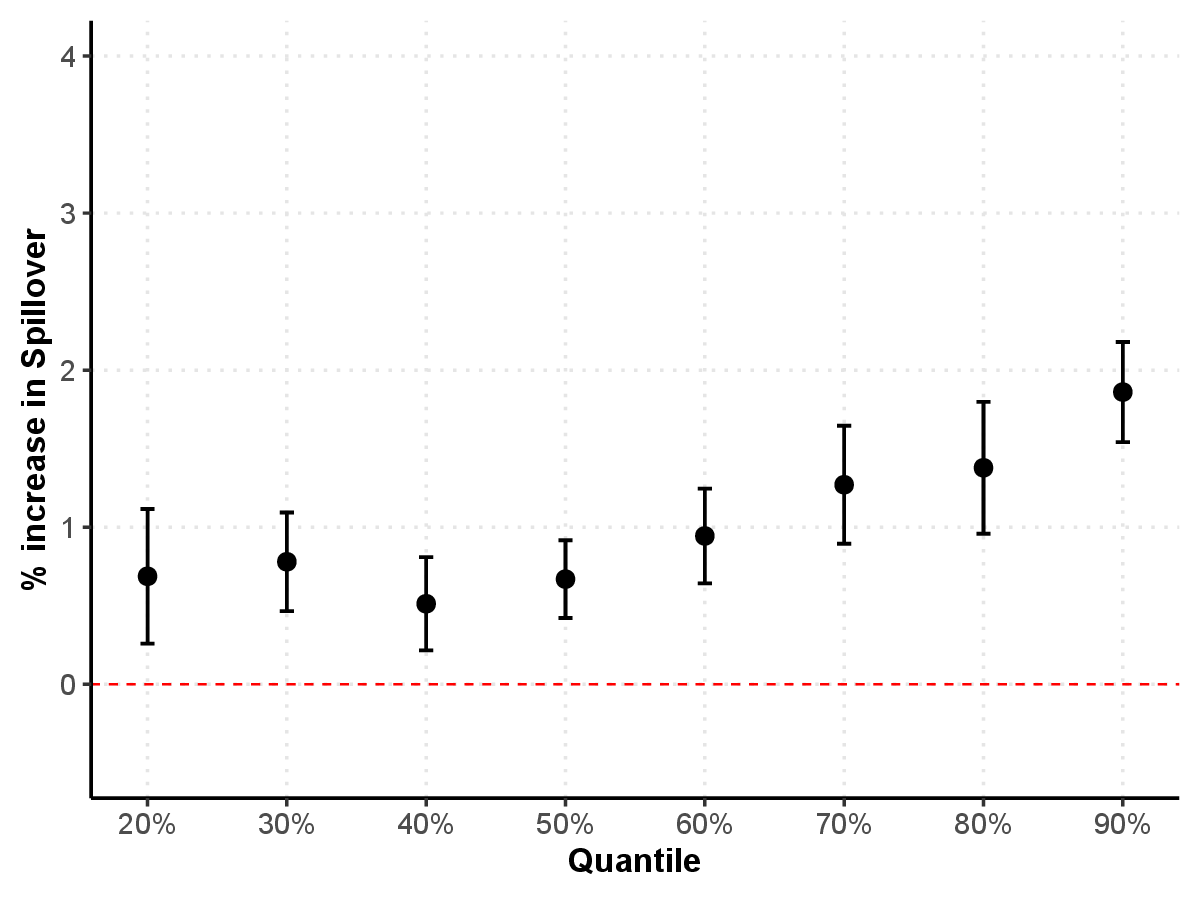}
        \caption{}
        \label{fig:qr_plot}
    \end{subfigure}
    \hfill
    \begin{subfigure}[b]{0.49\linewidth}
        \centering
        \includegraphics[width=\linewidth]{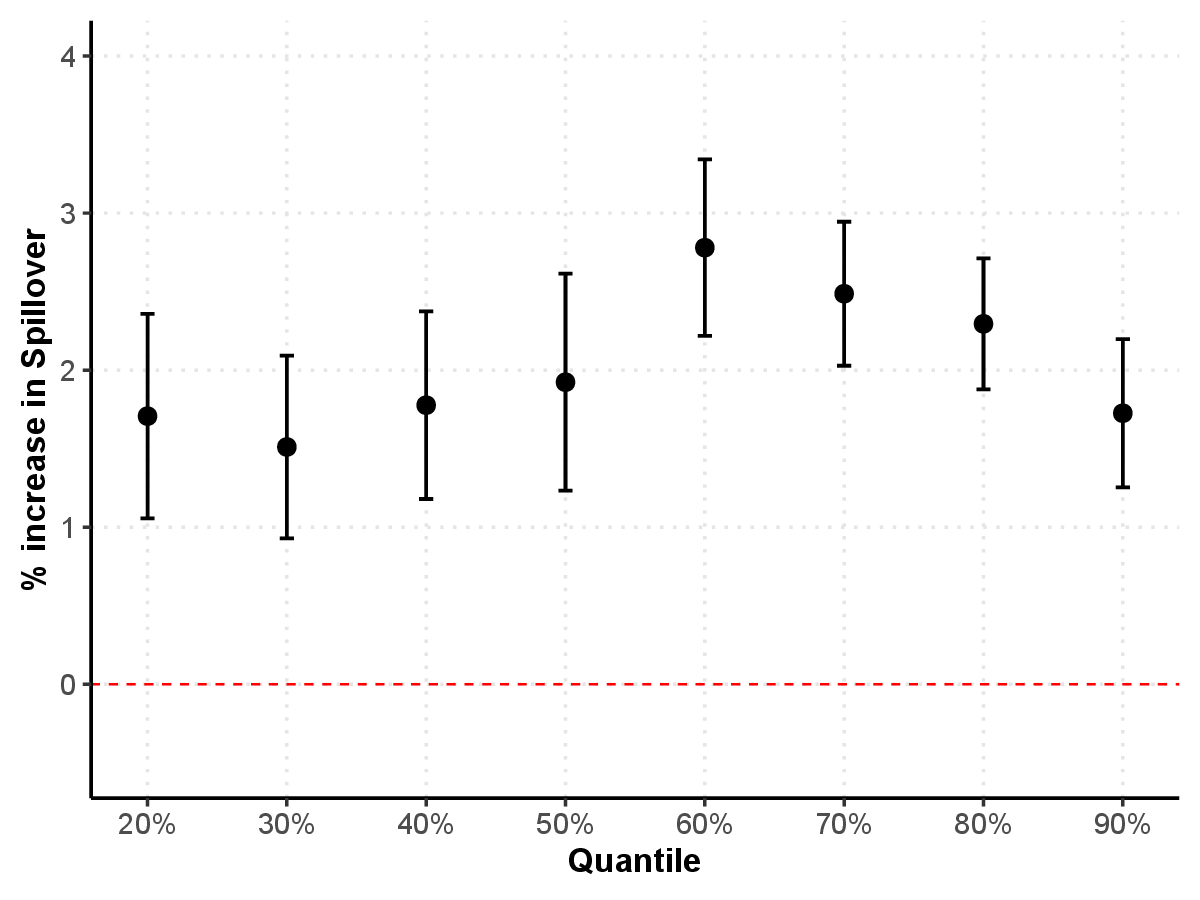}  
        \caption{}
        \label{fig:qr_plot_excl}
    \end{subfigure}
    \caption{Impact of a 100\% increase in non-zero conflict fatalities on spillover index percentiles (in percentage points) (a) Full sample relationship (b) Excluding 1628-1648}
    \label{fig:qr_plots}
\end{figure}

The raw data (Figure \ref{fig:scatterplots}) shows a positive and significant correlation between conflict and spillover ($r=0.17, p = 0.024$). However, this pattern breaks during the Thirty Years’ War, where spillover spiked early on, while fatalities remained constant throughout the conflict (by construction, as they are evenly spread). This is not surprising as it is unlikely that the model would maintain a high (over 3 standard deviations) level of spillover for 30 years. We therefore repeat the analyses keeping only the first decade of the conflict (excluding 1628-1648). When part of the Thirty Years’ War is excluded (Figure \ref{fig:scatterplot_excl}), the expected strong positive relationship becomes stronger ($r=0.42, p<0.001$). We confirm with an Ordinary Least Squares linear regression, controlling for the average CPI level in the region, in table \ref{tab:ols_models}. As an additional check,  we estimate the impact of a 1\% increase in non-zero conflict fatalities across the spillover distribution using quantile regressions, controlling for the average level of prices in the region again. The results in Table \ref{tab:qr_spillover} indicate a significant positive relationship across the distribution, even increasing at higher spillover percentiles: high level of spillover are explained by conflict fatalities. When the Thirty Years’ War is partly excluded (Table \ref{tab:qr_spillover_no30Y}), this relationship is both significant and consistent across the selected percentiles. Figure \ref{fig:qr_plots} displays the coefficients for additional percentiles, further supporting these results.

\begin{figure}[htbp!]
    \centering
    \begin{subfigure}[t]{0.5\linewidth}
        \centering
        \includegraphics[width=\linewidth]{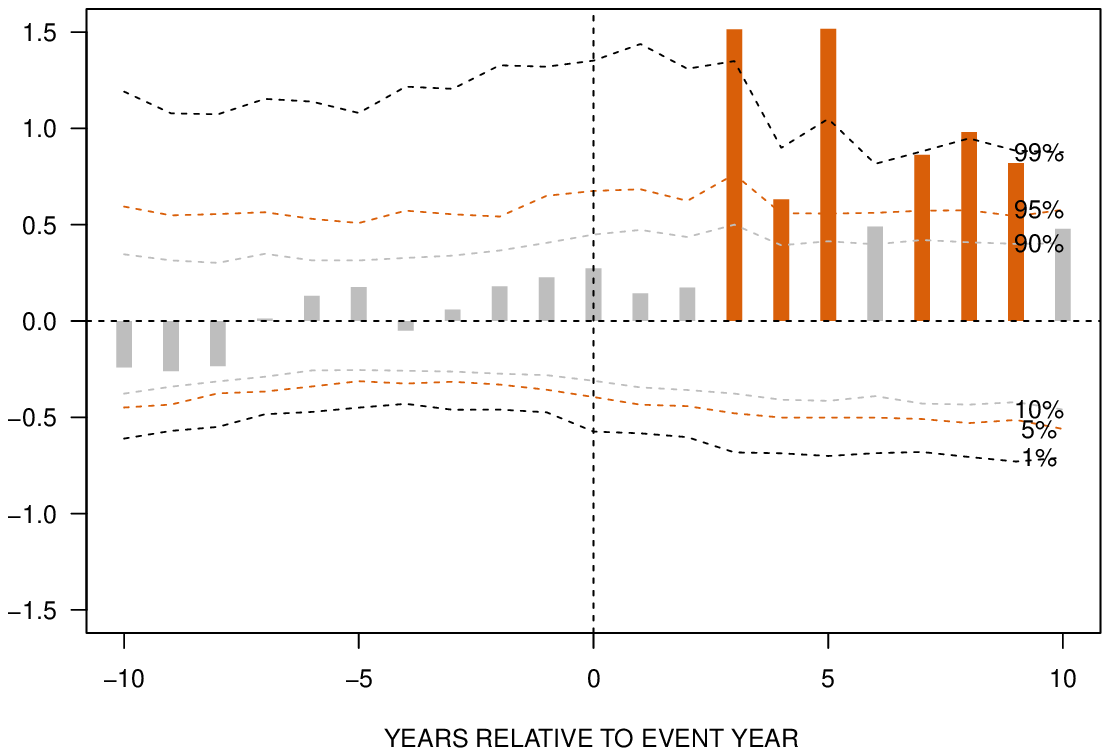}
        \caption{Start years}
        \label{fig:SEA_start}
    \end{subfigure}
    \hfill
    \begin{subfigure}[t]{0.5\linewidth}
        \centering
        \includegraphics[width=\linewidth]{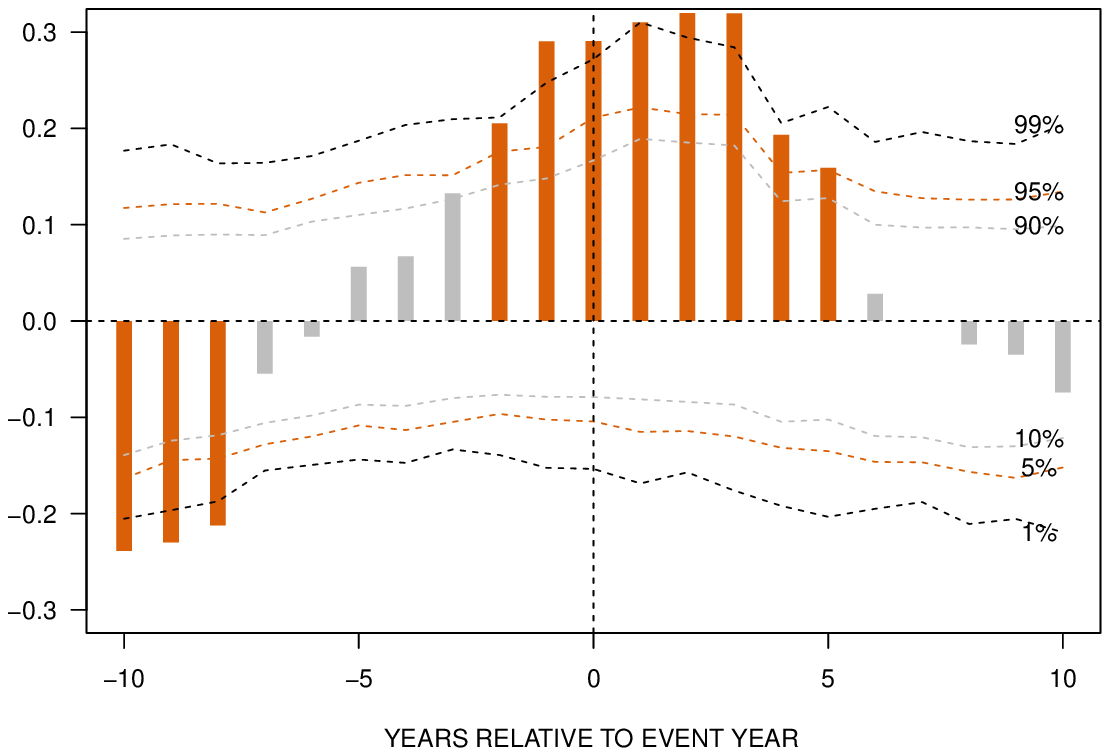}
        \caption{Entire period}
        \label{fig:SEA_all}
    \end{subfigure}
    \hfill
    \begin{subfigure}[t]{0.5\linewidth}
        \centering
        \includegraphics[width=\linewidth]{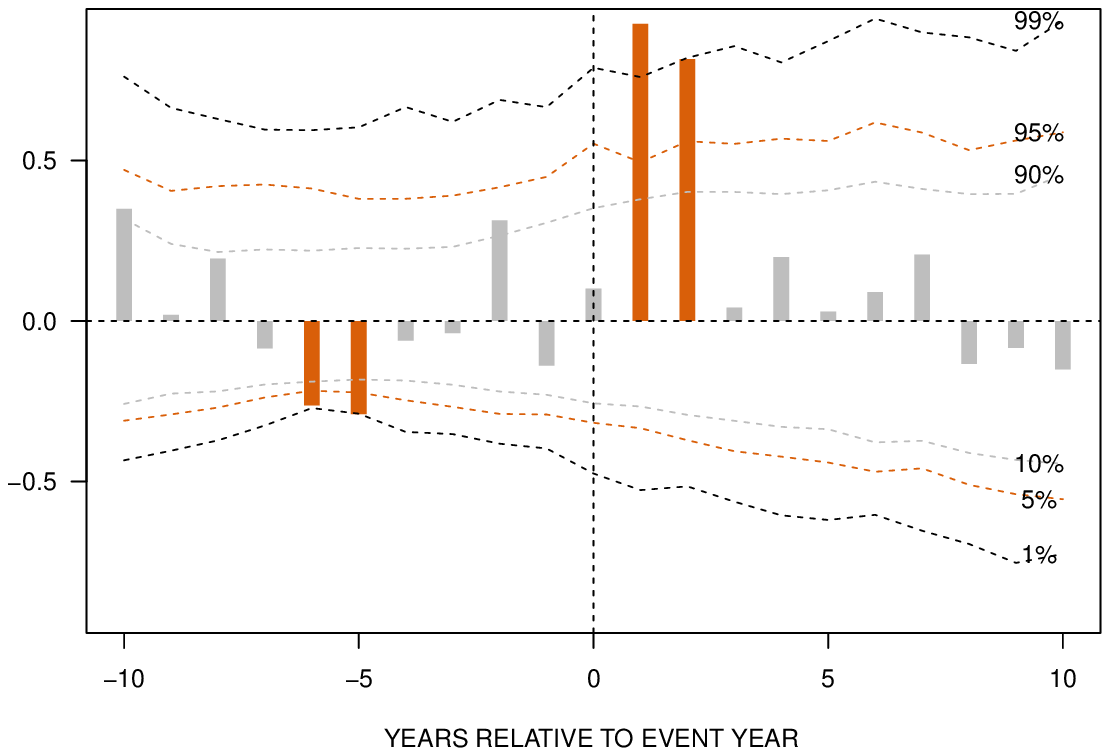}
        \caption{Midpoint}
        \label{fig:SEA_midpoint}
    \end{subfigure}
    \caption{Superposed Epoch Analysis using: (a) the start years of the four identified periods as events, (b) the entire period of these conflicts, and (c) the midpoint of these conflicts. Dashed lines represent statistical significance thresholds at indicated levels (1\%, 5\%, 10\%). Orange bars show significant deviations at the 5\% level.}
  
    \label{fig:SEA}
\end{figure}

Additionally, we run a Superposed Epoch Analysis (SEA), a statistical method used to identify the link between discrete events and continuous time or spatiotemporal processes and test the probability of such an association occurring by chance \cite{SEA81}. We run three SEAs using the time-normalization method of \citet{SEARao} and 10'000 bootstraps. The first SEA uses the set of years $\{1618,1688,1700,1701,1756\}$, corresponding to the start years of the identified key conflicts. The second one uses the entire set of years of these conflicts, except for the Thirty-Years War where we exclude 1628-1648 as we did previously. The third SEA uses the midpoint of each conflict (i.e. $
\text{midpoint} = \lceil \text{start} + \frac{\text{end} - \text{start}}{2} \rceil
$) which yields the following set of years $\{1633,1693,1708,1711,1759\}$. 
In all cases we study the average total spillover against the set of events. The results reveal a significant increase in spillover effects two years after the onset of conflicts (Figure \ref{fig:SEA_start}) and during the two years following the midpoint of the conflict (Figure \ref{fig:SEA_midpoint}). Furthermore, a notable rise in spillover effects is observed during periods centered around ongoing conflicts (Figure \ref{fig:SEA_all}).

\begin{table}
\caption{OLS regression of spillover index}
\begin{center}
\begin{threeparttable}
\begin{tabular}{l c c}
\hline
 &  (1) Full Sample & (2) Excl. 1628-1648 \\
\hline
(Intercept)     & $78.006 \; (3.997)^{***}$  & $65.563 \; (5.655)^{***}$ \\
log(Fatalities) & $0.754 \; (0.337)^{*}$     & $1.267 \; (0.397)^{**}$   \\
CPI             & $-13.858 \; (3.087)^{***}$ & $-7.075 \; (2.847)^{*}$   \\
\hline
N               & $168$                      & $147$                     \\
\hline
\end{tabular}
\begin{tablenotes}[flushleft]
\scriptsize{ \item $^{***}p<0.001$; $^{**}p<0.01$; $^{*}p<0.05$. \ Newey-West HAC standard errors are in parentheses. \ Dependent variable is the total spillover index (averaged over rolling windows of 30 to 40 years)}
\end{tablenotes}
\end{threeparttable}
\label{tab:ols_models}
\end{center}
\end{table}

\begin{table}
\caption{Quantile regression of spillover index}
\begin{center}
\begin{threeparttable}
\begin{tabular}{l c c c c}
\hline
 & 25th pctl & 50th pctl & 75th pctl & 90th pctl \\
\hline
Intercept       & $81.782 \; (2.256)^{***}$  & $81.767 \; (2.122)^{***}$  & $76.062 \; (5.467)^{***}$  & $67.386 \; (6.905)^{***}$ \\
log(Fatalities) & $0.559 \; (0.212)^{**}$    & $0.487 \; (0.185)^{**}$    & $1.028 \; (0.272)^{***}$   & $1.354 \; (0.258)^{***}$  \\
CPI             & $-17.331 \; (2.313)^{***}$ & $-15.000 \; (2.071)^{***}$ & $-13.062 \; (3.234)^{***}$ & $-6.270 \; (5.185)$       \\
\hline
N               & $168$                      & $168$                      & $168$                      & $168$                     \\
\hline
\end{tabular}
\begin{tablenotes}[flushleft]
\scriptsize{\item $^{***}p<0.001$; $^{**}p<0.01$; $^{*}p<0.05$. \ Bootstrapped standard errors are in parentheses. Dependent variable is the total spillover index (averaged over rolling windows of 30 to 40 years)}
\end{tablenotes}
\end{threeparttable}
\label{tab:qr_spillover}
\end{center}
\end{table}

\begin{table}
\caption{Quantile regression of spillover index (excl. 1628-1648)}
\begin{center}
\begin{threeparttable}
\begin{tabular}{l c c c c}
\hline
 & 25th pctl & 50th pctl & 75th pctl & 90th pctl \\
\hline
Intercept       & $73.803 \; (7.243)^{***}$ & $68.022 \; (5.634)^{***}$ & $60.647 \; (7.632)^{***}$ & $64.161 \; (10.439)^{***}$ \\
log(Fatalities) & $0.829 \; (0.379)^{*}$    & $1.293 \; (0.418)^{**}$   & $1.660 \; (0.398)^{***}$  & $1.343 \;  (0.449)^{**}$   \\
CPI             & $-12.523 \; (4.398)^{**}$ & $-9.695 \; (2.663)^{***}$ & $-4.387 \; (4.009)$       & $-3.051 \;  (6.530)$       \\
\hline
N               & $147$                     & $147$                     & $147$                     & $147$                      \\
\hline
\end{tabular}
\begin{tablenotes}[flushleft]
\scriptsize{\item $^{***}p<0.001$; $^{**}p<0.01$; $^{*}p<0.05$. \ Bootstrapped standard errors are in parentheses. Dependent variable is the total spillover index (averaged over rolling windows of 30 to 40 years)}
\end{tablenotes}
\end{threeparttable}
\label{tab:qr_spillover_no30Y}
\end{center}
\end{table}


Our findings support that warfare significantly disrupted contagion dynamics and has led to spillover increases. The four highlighted periods of extreme conflict are statistically linked with price disruptions across markets, emphasizing their interconnectedness rather than their independence. 

\subsubsection{Network propagation effects}
\begin{figure}[htbp!]
    \centering
    \includegraphics[width=0.99\linewidth]{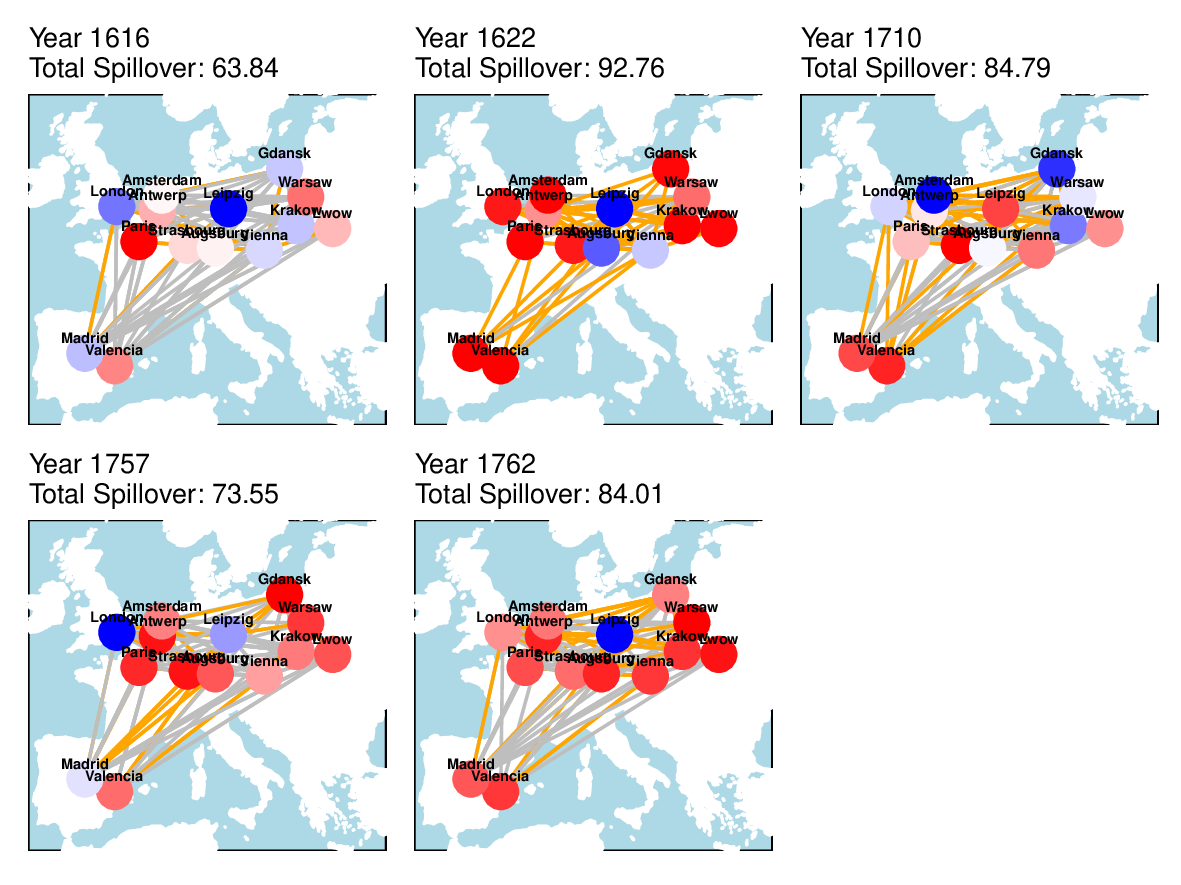}
    \caption{Evolution of the contagion network. The color of the node represents net spillover flows (blue is net risk transmission, red is net risk reception). Edges with spillover $>$ 10 are in orange. Periods highlighted are (from top left to bottom right): 1. Before 30-Years War, 2. 30-Years War, 3. Great Northern War and War of Spanish Succession, 4 and 5. Seven Years War }
    \label{fig:contagion_network_evolution}
\end{figure}

Estimated spillovers or general statistical relationships should not be interpreted as definite causal claims in our opinion. Instead, we view these quantitative methods as tools to guide deeper exploration into qualitative literature, to assess whether model-based statistical relationships are causal. To better understand the links between price disruptions and warfare, we study the evolution of the network representation of the spillover matrix (seen figure \ref{fig:contagion_network_evolution}), as well as the pairwise spillovers (seen in figure \ref{fig:net_spillover_figs}). 

\begin{itemize}
    \item During the Thirty-Years War, the Holy Roman Empire was the main spillover transmitter, with peak contagion around 1622. We believe this accurately captures the crisis of 1618-1623 in Germany, a known period of hyperinflation referred to as "Kipper und Wipper", coherent with the fact that the War was primarily fought in Germany, leading to price shocks that propagated in accordance with Gresham's Law (bad money drives out good money) \cite{germancrisis16181623, ogilvie1992germany}. Our results suggest that this propagation was Europe-wide.

   \item During the Great Northern War and the War of Spanish Succession, we see most polish cities emitting risk, a non-standard behavior when looking at the long run dynamics and the pre and post Thirty-Years' War network maps. We see a likely confounder in the Plague outbreak that erupted around 1709, which severely affected Polish cities \cite{plaguebaltic, plaguesnature}.

   \item Finally during the Seven Years War, we note two main trends. First, London transmitted spillover in the early years of the war (figure \ref{fig:net_spillover_city}), which was then compounded by Leipzig around 1762. (figure \ref{fig:contagion_network_evolution}).
   
The spillover we see coming from London might be due to dual effects of the War and the beginning of the Industrial Revolution in the early 1760s, which propelled London as an international trade hub, disrupting the trade dynamics with its neighboring partners Paris, Amsterdam, Antwerp (figure \ref{fig:net_spillover_city}). The British economy suffered through a heavy debt burden used to finance the War \cite{oxfordsevenyears},and has drawn substantially upon extra-national sources of capital, labor and materials\cite{7yrs_casestudy}, which might explain the source of spillover as well.

The spillover coming from Leipzig in the later stages of the War is more intriguing. We do see a price spike in the raw data but the model tells us this spike severely affected other markets, and precisely mostly Polish cities (see Roman Empire and Poland in \ref{fig:net_spillover_country}. We see two convincing overlapping factors at play here: the first one would be the battle of Torgau (near Leipzig) in 1760, one of the bloodiest conflicts of the Third Salesian War with close to 15'000 casualties \cite{duffy1985frederick}, and a general instability during Prussian occupation in 1759-1760: Leipzig was occupied by Prussians, who the withdrew in August 1759, recaptured in September, and withdrew again in October 1760 \cite{szabo2013seven}. The second factor is that Polish cities were very affected by turmoil in Leipzig because of their inherent fragility in the wake of the bloody (well over 15'000 deaths\cite{redman2014frederick}) Battle of Zorndorf in 1758 (now Sarbinowo, West of Gdansk on the coast, in Poland).
    
\end{itemize}

\subsection{Spillover Dynamics between main economic trade centers}

We first discuss the long-term behavior of the studied markets and then turn to a more detailed dynamic analysis to understand the shifts in relationships between different cities. We expect to find that the main economic powerhouses of the early-modern period were the most influential markets, and biggest sources of contagion. 
\subsubsection{Full Sample Analysis}

\begin{figure}[htbp!]
    \centering
    \begin{subfigure}[b]{0.49\linewidth}
        \centering
        \includegraphics[width=\linewidth]{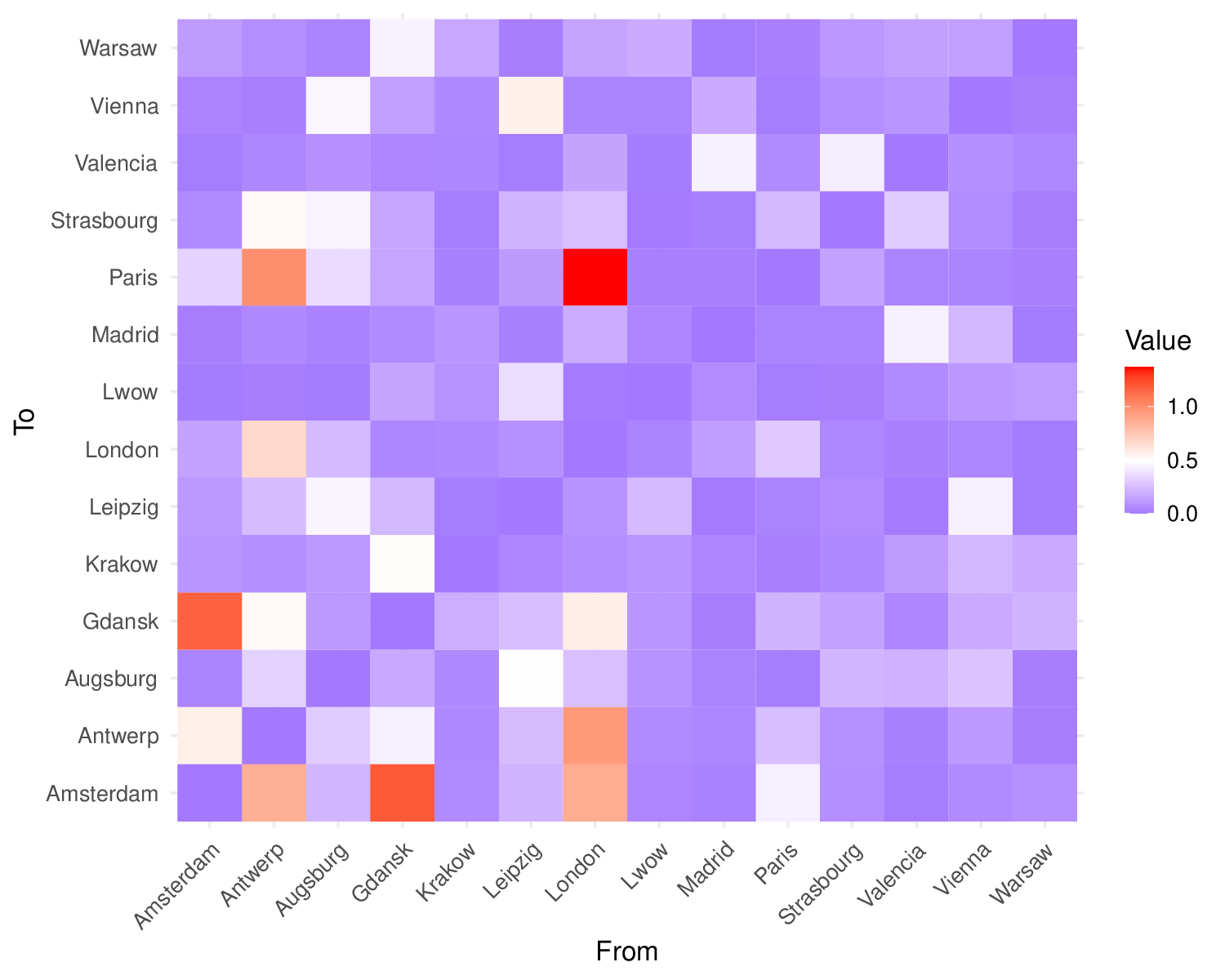}
        \caption{}
        \label{fig:DYTable}
    \end{subfigure}
    \hfill
    \begin{subfigure}[b]{0.49\linewidth}
        \centering
        \includegraphics[width=\linewidth]{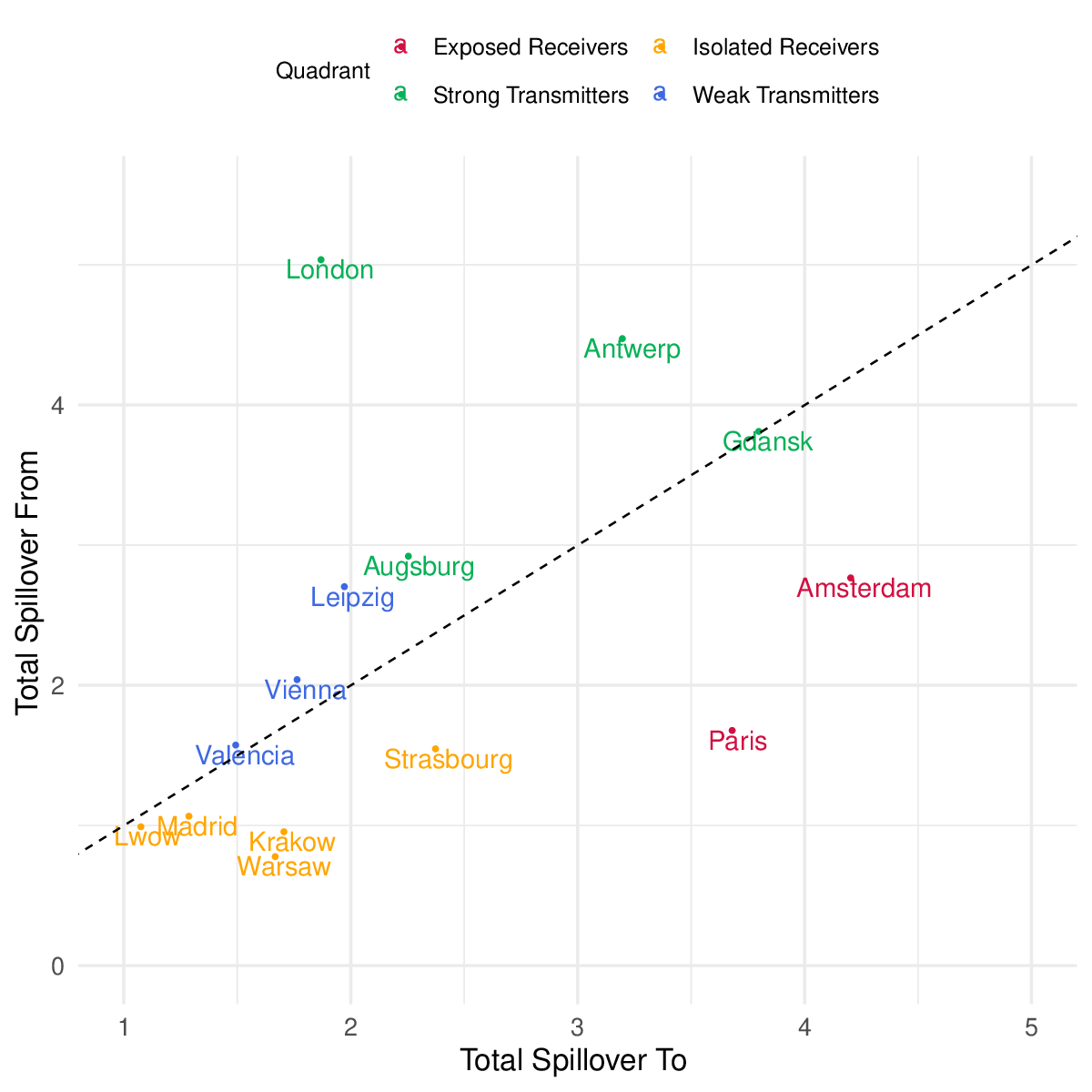}  
        \caption{}
        \label{fig:DYscatter}
    \end{subfigure}
    \caption{ (a) Heatmap of the full-sample Spillover Table, standardized (b) Scatter plot with x-axis representing total inward "to" spillover, and y-axis representing total outward "from" spillover. Transmitter designates positive net spillover. Strong Transmitter have outwards spillover above 75th percentile and Weak Receivers have inwards spillover above 75th percentile. }
    \label{fig:DYsummary}
\end{figure}

\begin{figure}[htbp!]
    \centering
    \begin{subfigure}[b]{0.45\linewidth}
        \centering
        \includegraphics[width=\linewidth]{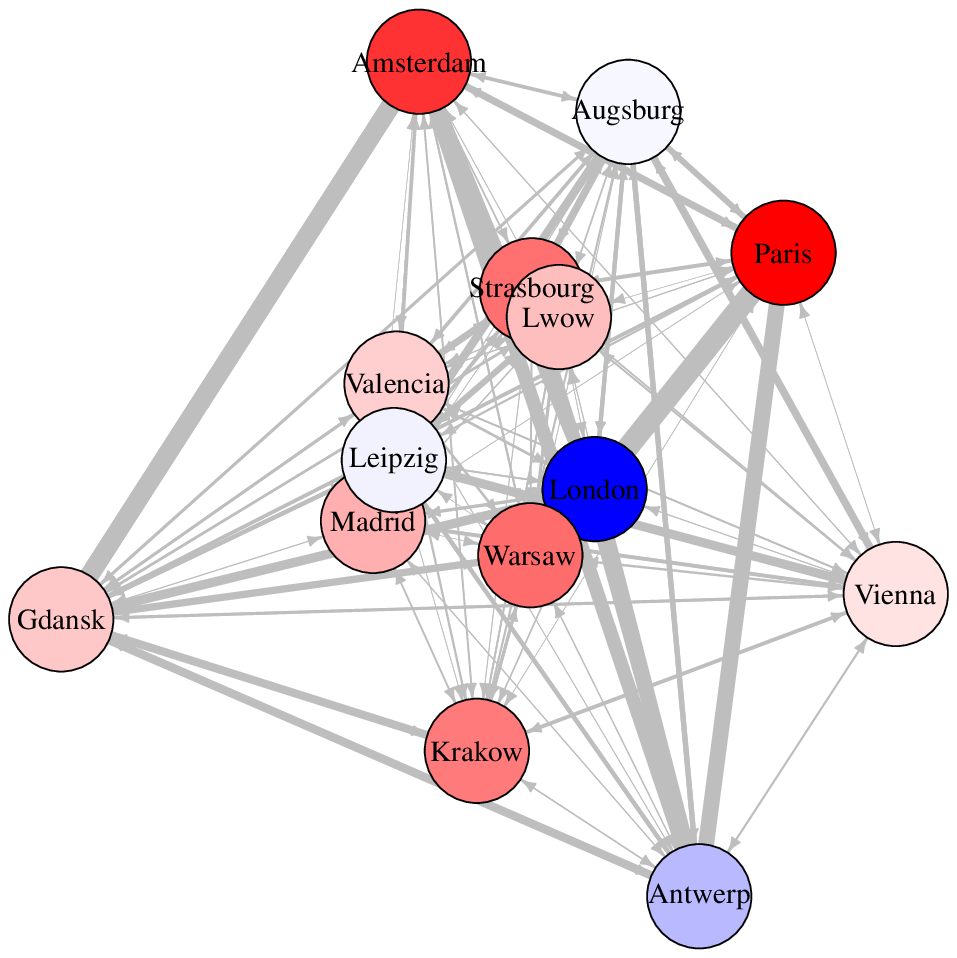}
        \caption{}
        \label{fig:DYNet}
    \end{subfigure}
    \hfill
    \begin{subfigure}[b]{0.47\linewidth}
        \centering
        \includegraphics[width=\linewidth]{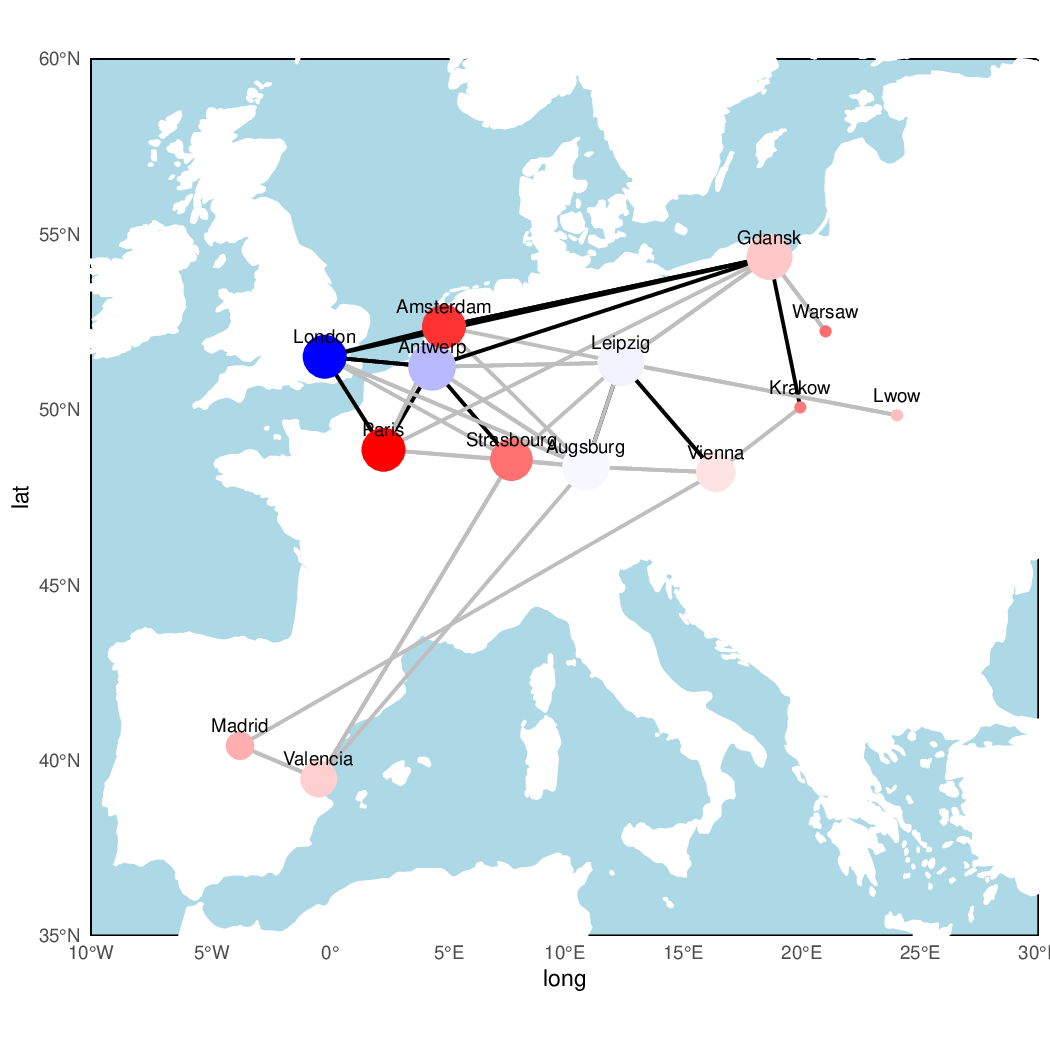}  
        \caption{}
        \label{fig:DYNetmap}
    \end{subfigure}
    \caption{Graph representations of the full-sample spillover table. Colors indicate the net spillover for the location: blue is positive net spillover (transmitter), red is negative net spillover (receiver) (a) Network Graph with fully connected nodes. Edge width is proportional to $w_{ij}$ the weight on the adjacency matrix, i.e. the spillover from $i$ to $j$ (b) Spatial representation of the network graph. We only retain edge weights $w_{ij} >0.2$ and color high spillover edges $w_{ij} > 0.5$ in black. The node size is given by $\log(w_{ij})$  }
    \label{fig:DYnets}
\end{figure}

We first assess the long-run average dynamics of the model through the spillover matrix and its network graph representation, figures \ref{fig:DYsummary} and \ref{fig:DYnets}. While we defer to the dynamic analysis for more historically nuanced analysis, as looking at average behavior for a 200 year period has limitations in failing to capture interesting but shorter-run dynamics, we do see historical coherence in the results.\\
As we expected, the main trade centers of the early modern period century, Amsterdam, Antwerp, Gdansk and London are among the highest emitters or receivers of spillover.

After 1585 the dominant forces are the Dutch conurbation of Amsterdam with Genoa, and economic power shifts decidedly towards Amsterdam after 1600. There is strong bi-directional spillover between Gdansk and Amsterdam, in agreement with their strong trade relationship throughout the early-modern period \cite{heeres1988dunkirk}, as Gdansk played a major role in the rise of Amsterdam as its main supplier of grain.

Antwerp's decline after its Fall in 1583 following the Schelde's closure  reinforced Amsterdam's position as the main European mercantile hub \cite{de1997dutchecon} and explains the net spillover direction (Antwerp $\rightarrow$ Amsterdam).  Amsterdam is the last of this series of economically dominant central staplemarkets. 

The analysis characterizes London as an active trade node. While the capital was not a powerhouse in the earlier part of the early modern period (1600s) by the early 1700s, according to \citet{ormrod2003rise}, international trade shifted to London, which was not a staplemarket but rather a metropolis with an integrated national economy as hinterland and built its success on the combination of a strong nation state with a colonial empire.

\begin{figure}[htbp!]
    \centering
    \begin{subfigure}[b]{0.8\linewidth}
        \centering
        \includegraphics[width=\linewidth]{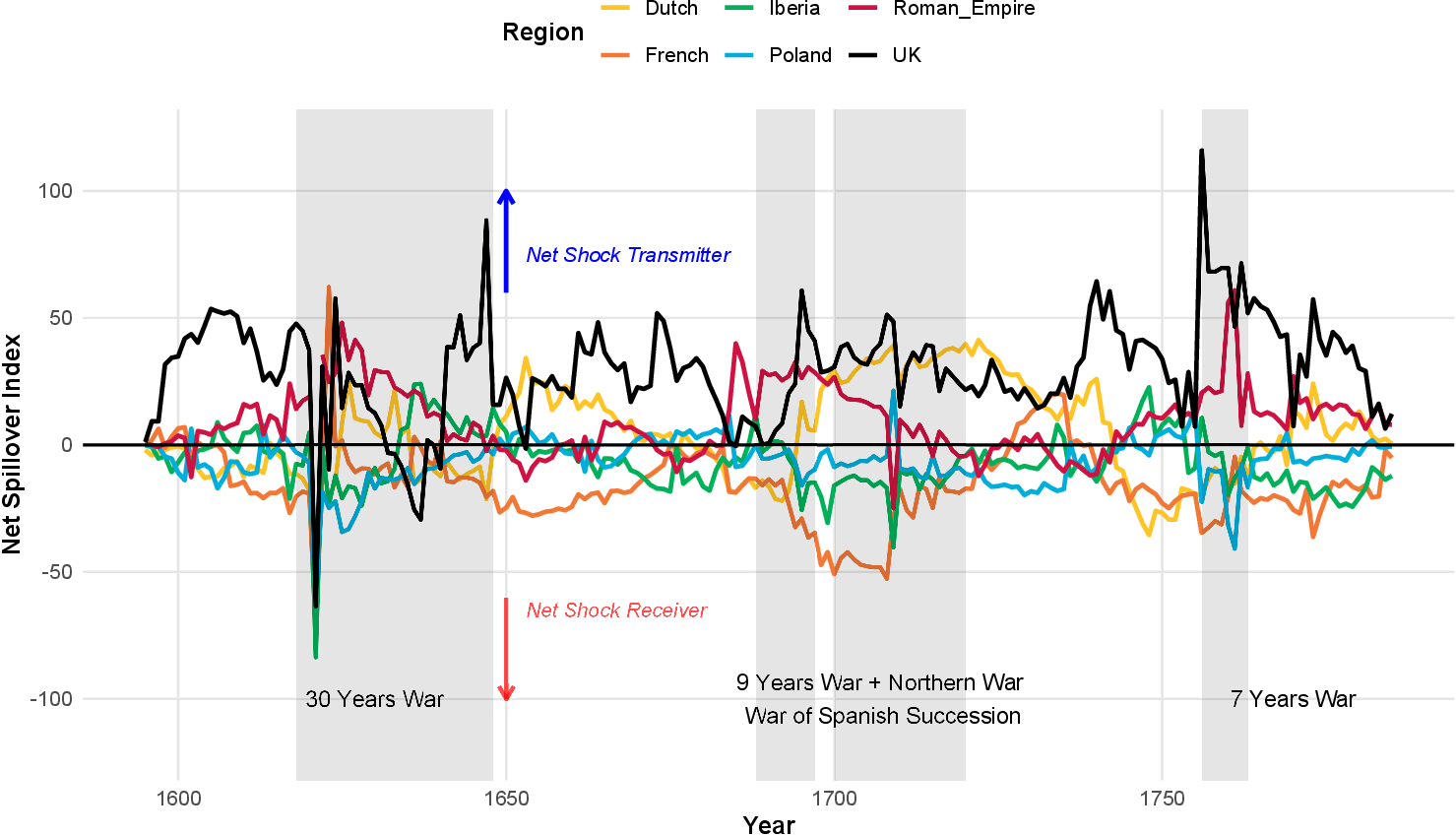}
        \caption{}
        \label{fig:net_spillover_country}
    \end{subfigure}
    \hfill
    \begin{subfigure}[b]{0.8\linewidth}
        \centering
        \includegraphics[width=\linewidth]{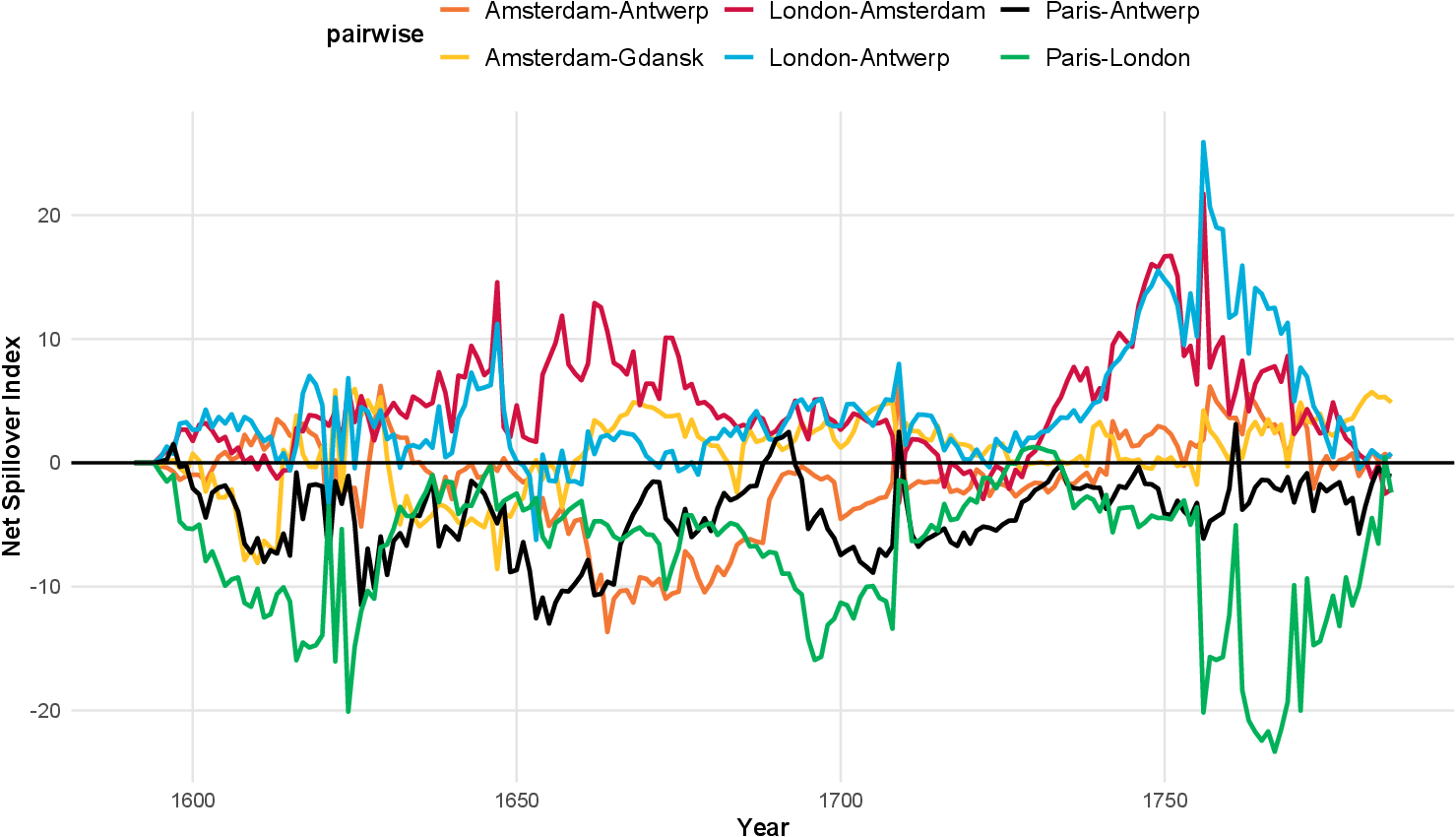}  
        \caption{}
        \label{fig:net_spillover_city}
    \end{subfigure}
    \caption{ (a): Net Spillover aggregated by political/regional affiliation. A positive net spillover means the region emits risk, while a negative net spillover means the region is receiving risk (b) Net Pairwise spillovers. Naming convention 'A-B' is spillover from A to B. }
    \label{fig:net_spillover_figs}
\end{figure}

We will now turn to analyzing the contagion patterns between the four most emblematic economic centers of the panel: London, Amsterdam, Paris, Antwerp. We supplement long run relationships of the previous section with the dynamic net pairwise spillovers of our locations of interest plotted in figure \ref{fig:net_spillover_figs}.

We begin by addressing the role of London as a significant risk transmitter within the network. Our analysis reveals substantial spillovers originating from London to numerous locations, particularly during the Seven Years' War and, even more notably, the Industrial Revolution in the 1760s. Both phenomena were addressed in the previous section. 

Even when accounting for those factors, there remains significant spillover from London to Paris in both static and dynamic analyses. Notably, spillover levels were elevated and steadily increasing from the early 1600s through the 1630s. We will not address those as most historical price reconstructions, icluding Allen's, fail to capture . Similarly, high spillover is observed in the 1690s, a period marked by the Nine Years' War. As documented by \citet{fierro1996histoire}, the war brought increased taxation for Parisians, and poor harvests between 1692 and 1694 caused widespread hunger in the city. These factors may explain why Paris was more vulnerable to spillover effects from London compared to other locations during this era.
    
Amsterdam was also notably vulnerable to spillover from its neighbors Antwerp and London in the second half of the 1660 to 1680s. That specific period contained many events to justify increased shock propagation between these three cities: the 1670s marked the end of the Dutch Golden Age with Rampjaar (the "Disaster Year") in 1672, during the Franco-Dutch War. This decade was indeed one of constant warfare for the Dutch, with the Anglo-Dutch War and the Franco-Dutch War. We also note a substantial plague outbreak in Amsterdam from 1663 to 1666 which made the city more vulnerable to spillover from its neighbors.


\section{Discussion}

\subsection{Interpretation Challenges}

We first address conceptual questions on the causal nature of the relationships identified by the model. The spillover is a statistical estimate of the sensitivity of one variable to another, which we view as different from a causal claim of whether the change in one variable caused the change in the other. We see the quantitative information given by the model as a starting point to investigate potentially causal relationships using qualitative literature to cross-reference our findings and ground the results in historical precedent.

In our study, the model demonstrates a high level of coherence: it accurately maps out the spillover network, identifying the primary mercantile cities of early modern Europe as key nodes. Additionally, it captures specific, temporally and spatially localized events, such as the spillover effects from Poland due to the 1709 plague and from Germany during the 1618-1623 crisis.

On the other hand, the apparent price shock propagation of London on Paris in the early 1600s to 1630 is puzzling. We attribute this part of the results to data deficiencies in Paris prices during the 1600-1630 period (see appendix for further details).

\subsection{Methodological Limitations }

The first limitation of this analysis is its natural reliance on the data used. We deliberately chose Bob Allen's CPI measure as it has become a reference dataset in the past 20 years, widely used by economic historians \cite{allen2001_ref1, allen2001_ref2, allen2001_ref3, allen2001_ref4}. We also opted for a single source of data to preserve methodological consistency across locations and time periods.  Naturally, a study of trade dynamics instead of price dynamics could provide more concrete and interpretable insights in this specific context of shock propagation, but such an analysis is currently constrained by the scarcity of relevant data.

On the computational side, a major challenge in historical applications of this model is the "curse of dimensionality". Most historical datasets with multi-location price data are often rank-deficient or too limited to support a high-order VAR model, effectively making the dynamic spillover index impossible to estimate. One potential solution to this issue is the use of sparse VAR models \cite{ReinselRRVAR,nicholson2017varxlstructuredregularizationlarge,nicholson2020high}, although this approach introduces the need for hyper-parameter selection. To obtain a robust measure of spillover, one would then need to compute a generalized forecast-error variance decomposition (FEVD),made invariant to variable ordering \cite{kpps1996impulse,kppspesaran1998generalized,lanne2016GFEVD}. 

A promising direction for future research would be the incorporation of exogenous shocks into the model. Many applications would benefit from being able to study the impact of an outside variable on the shock propagation dynamics of the system of variables This, however, presents a methodological challenge: estimating the FEVD for VAR models with exogenous variables (VARX), where shocks arise from both dependent and independent variables, poses significant identification issues.

\section{Conclusion}

We have explored the dynamics of market integration and price contagion in early modern Europe, focusing on how significant events such as wars and political upheavals influenced economic interdependencies among European cities. By applying the Diebold-Yilmaz contagion model, we have provided a novel quantitative assessment of how economic shocks propagated across different markets during this period.

Our analysis reveals that warfare was a critical driver of economic contagion. We find that severe conflicts increased price shock propagation, suggesting that markets were sensitive to each other and not isolated in times of systemic stress. Notably, during the Thirty Years' War, the Holy Roman Empire emerged as a major transmitter of economic shocks, which aligns with historical accounts of hyperinflation and monetary instability during the period. Similarly, the Seven Years' War saw London emerge as a central node of risk transmission in the 1760s, reflecting its economic nascent prominence as well as the British Empire's heavy debt burden to finance the Seven Years War.

We believe this research contributes to the understanding of market integration by providing empirical evidence of the mechanisms through which economic shocks were transmitted across regions. The application of the Diebold-Yilmaz framework, typically used in contemporary financial and trade network studies, demonstrates its utility in historical economic analysis, offering new insights into the economic underpinnings of Europe’s historical development. Future studies could build on this approach by incorporating more granular data and exploring the interplay between economic and non-economic factors in shaping market integration and economic resilience in historical contexts.

\bibliographystyle{unsrtnat}
\bibliography{references.bib}

\section{Appendix}

\subsection{Paris CPI data deficiency in 1600-1629}

We consider that Allen's CPI data are unreliable for the 1600-1629 period.
after checking the actual data series in \citet{hauser1985recherches} and \citet{hanauer1876etudes}.
 We checked data availability against the weights that Allen used to calculate the CPI. While London has 79.4\% of the required data, Paris has only 39.7\%, with the remainder derived from Strasbourg data. Prices for eggs seem to have been derived from outside London and doubled to calculate London prices.

As Allen states, there are many gaps in the Paris series: for the 30 years 1600-1629, we have only 7 data points for fuel (wood) and 2 data points for butter. Some of the commodities are not comparable, such as beer and wine, or charcoal and wood. Bread prices for both London and Paris were computed from wheat prices, which is not a problem in itself but has implications for comparison with other cities. See table \ref{tab:cpi_weights} for a detailed comparison.

\begin{table}[ht!]
\centering
\caption{CPI Weights and Data Availability, Allen (2001, p.421)}
\label{tab:cpi_weights}
\begin{tabular}{l>{\raggedright}p{2cm}>{\raggedright}p{2cm}>{\raggedright\arraybackslash}p{2cm}}
\toprule
Commodity & CPI Weight (\%) & London (\% Data Availability) & Paris (\% Data Availability) \\
\midrule
Bread & 30.4 & 30.4 & 30.4 \\
Beans/Peas & 6.0 & 6.0 & 6.0 \\
Meat & 13.9 & 13.9 & 13.9 \\
Butter & 4.3 & 4.3 & 4.3 \\
Cheese & 3.6 & 3.6 & 3.6 \\
Eggs & 1.3 & 1.3 & 1.3 \\
Beer/Wine & 20.6 & 0.0 & 20.6 \\
Soap & 1.8 & 1.8 & 1.8 \\
Linen & 5.3 & 5.3 & 0.0 \\
Candles & 3.1 & 3.1 & 3.1 \\
Lamp Oil & 4.7 & 4.7 & 4.7 \\
Fuel (Charcoal/Wood) & 5.0 & 5.0 & 5.0 \\
\midrule
\textbf{Total} & \textbf{100.0} & \textbf{79.4} & \textbf{39.7} \\
\bottomrule
\end{tabular}
\end{table}

Paris had a large aristocratic populace which accepted very high prices for conspicuous consumption. The Paris market was much more expensive than coastal French cities due to transport costs and aristocratic consumers. London, on the other hand, benefited from easier transport. It is therefore important to know if Allen’s Paris prices represent market prices for everyday consumption or the much higher prices negotiated for high-quality products contracted for the elite (see \citet{allaireholm2022}). Unfortunately, \citet{hauser1985recherches} simply states: “tous les prix sont tirés de divers documents des Archives Nationales” (all prices are from various National Archives documents). So the price data may reflect very different social contexts.

\end{document}